\documentclass[sigconf]{acmart}
\AtBeginDocument{%
  \providecommand\BibTeX{{%
    \normalfont B\kern-0.5em{\scshape i\kern-0.25em b}\kern-0.8em\TeX}}}

\setcopyright{acmcopyright}
\copyrightyear{2023}
\acmYear{2023}
\setcopyright{acmlicensed}\acmConference[CHI '23]{Proceedings of the 2023 CHI Conference on Human Factors in Computing Systems}{April 23--28, 2023}{Hamburg, Germany}
\acmBooktitle{Proceedings of the 2023 CHI Conference on Human Factors in Computing Systems (CHI '23), April 23--28, 2023, Hamburg, Germany}
\acmPrice{15.00}
\acmDOI{10.1145/3544548.3580791}
\acmISBN{978-1-4503-9421-5/23/04}

\usepackage{enumitem}
\usepackage{multirow}
\usepackage{subcaption}
\usepackage{soul}
\usepackage{xcolor}

\definecolor{highlighter}{HTML}{fff100}
\sethlcolor{highlighter}

\newcommand{\JS}[1]{}
\newcommand{\DP}[1]{#1}
\newcommand{\submission}[1]{}
\newcommand{\remove}[1]{}
\newcommand{\revision}[1]{#1}

\begin{document}

\title[Don't Look at the Data!]{Don't Look at the Data! How Differential Privacy Reconfigures the Practices of Data Science}

\author{Jayshree Sarathy}
\authornote{Corresponding author}
\email{jsarathy@g.harvard.edu}
\affiliation{%
  \institution{Harvard John A. Paulson School for Engineering and Applied Sciences and OpenDP}
  \streetaddress{150 Western Ave}
  \city{Boston}
  \state{MA}
  \postcode{02134}
  \country{USA}
}

\author{Sophia Song}
\authornote{This work was done when the author was an intern with the Privacy Tools Project at Harvard University, supported by the REU program.}
\affiliation{%
  \institution{UC Berkeley}
  \city{Berkeley}
  \state{CA}
  \country{USA}
}

\author{Audrey Haque}
\affiliation{%
  \institution{Harvard John A. Paulson School for Engineering and Applied Sciences, Harvard Graduate School of Design, and OpenDP}
    \streetaddress{150 Western Ave}
  \city{Boston}
  \state{MA}
  \postcode{02134}
  \country{USA}
}

\author{Tania Schlatter}
\affiliation{%
 \institution{Harvard Institute for Quantitative Social Sciences and OpenDP}
 \city{Cambridge}
 \state{MA}
 \country{USA}
 }

\author{Salil Vadhan}
\affiliation{%
  \institution{Harvard John A. Paulson School for Engineering and Applied Sciences and OpenDP}
    \streetaddress{150 Western Ave}
  \city{Boston}
  \state{MA}
  \postcode{02134}
  \country{USA}
 }

\renewcommand{\shortauthors}{Sarathy et al.}

\begin{abstract}
 
Across academia, government, and industry, data stewards are facing increasing pressure to make datasets more openly accessible for researchers while also protecting the privacy of data subjects. Differential privacy (DP) is one promising way to offer privacy along with open access, but further inquiry is needed into the tensions between DP and data science. In this study, we conduct interviews with 19 data practitioners who are non-experts in DP as they use a DP data analysis prototype to release privacy-preserving statistics about sensitive data, in order to understand perceptions, challenges, and opportunities around using DP. We find that while DP is promising for providing wider access to sensitive datasets, it also introduces challenges into every stage of the data science workflow. We identify ethics and governance questions that arise when socializing data scientists around new privacy constraints and offer suggestions to better integrate DP and data science. 
\end{abstract}

\ccsdesc[500]{Security and privacy~Social aspects of security and privacy}
\ccsdesc[500]{Security and privacy~Usability in security and privacy}
\ccsdesc[300]{Security and privacy~Privacy-preserving protocols}

\keywords{privacy, utility, open access, data practitioners, data analysis}


\maketitle

\section{Introduction}

Researchers, government agencies, and companies are increasingly expected to share their datasets with 
other researchers and 
the public~\cite{burwell2013open,gherghina2013data,vlaeminck2015data}.
Data repositories such as Dataverse~\cite{magazine2011dataverse,king2007introduction} and Dryad~\cite{white2008dryad}
exist to promote such data sharing by ingesting and preserving datasets to be shared in the long-term.
Yet, many datasets contain sensitive information about individuals.
Over the last two decades, increases in computational power and availability of data sources have enabled new threats to the privacy of sensitive datasets,
and it has been shown that heuristic anonymization techniques, such as removing personally identifiable information or only releasing aggregate statistics, do not adequately protect privacy ~\cite{narayanan2008robust,dwork2017exposed}.
Growing calls for open access along with growing threats to privacy mean that data stewards are caught in a bind: either they release data and potentially compromise privacy, or they must put in place restrictive and costly mechanisms  before allowing researchers access to their datasets.

To ease this tension, many organizations are turning to formal privacy frameworks such as differential privacy (DP) ~\cite{dwork2006calibrating} to protect privacy while offering wider access to rich datasets. But even as DP has gained prominence through high-profile deployments at Google~\cite{erlingsson2014rappor}, Apple~\cite{greenberg2016apple}, and the U.S. Census Bureau~\cite{abowd2018us,machanavajjhala2008privacy}, these deployments have also illuminated the significant challenges in bringing privacy-preserving data science from theory to practice.  
Scholars have begun to highlight the tensions between DP and data science, such as differing conceptions of risk, and ways in which DP data analysis clashes with ingrained workflows and modes of interaction within statistical agencies ~\cite{drechsler2021differential,oberski2020differential}. 
However, these scholars also point out that DP may lead to better scientific research by protecting against p-hacking and introducing robustness into the data analysis process~\cite{dwork2009differential,dwork2015reusable,oberski2020differential}. 
Others \remove{draw attention to}\revision{foreground} the ``scant attention given to socialization of [privacy] tools” ~\cite{gurses2016privacy} and epistemic disconnects about data exposed by DP ~\cite{boyd2022differential} as obstacles for bringing privacy-preserving methods from the lab into the world.
These works highlight critical gaps between theory and practice of privacy-preserving data science that are crucial to explore further. 


In this work, we offer insight into opportunities and challenges of DP from the perspective of data analysts, depositors, and administrators.\footnote{See Section~\ref{sec:dp-creator} for a description of these roles, and Section~\ref{sec:methods} for more details on our participant sample.} 
In the HCI community, studies have started to explore challenges for design, communication, and governance of privacy-preserving data analysis tools, but they have largely done so through user studies examining the perspectives of 
data subjects~\cite{cummings2021need,xiong2020towards,bullek2017towards,smart2022understanding} and developers of DP software~\cite{agrawal2021exploring}. Less studied are the needs of \textit{data practitioners}, such as data depositors, administrators, and analysts, who are non-experts in DP with varying technical backgrounds, and who are experienced in sharing and analyzing sensitive datasets\revision{, particularly in the context of research data repositories}. 
In addition, we consider the utility of differentially private data analysis tools towards the broader aims often stated for the use of DP, including exploratory data analysis, replication of scientific studies, and wider access to the public~\cite{gaboardi2016psi}.

In particular, we consider the following research questions.
\begin{itemize}[label={}]
    \item \textbf{RQ1}: What barriers do data practitioners who are non-experts in DP face when using DP to share or analyze sensitive datasets?
    \remove{\item \textbf{RQ2}: What is the potential utility of DP for advancing goals of safe, open access to sensitive datasets?}
    \item \revision{\textbf{RQ2}: What do data practitioners who are non-experts in DP perceive to be the potential utility of DP for expanding access of sensitive data to the public, facilitating exploratory data analysis, and enabling replication of scientific studies?}	
    \remove{\item \textbf{RQ3}: What are the broader implications of DP for data science?}
    \item \revision{\textbf{RQ3}: What changes need to be made in the data science workflow in order to address the barriers and realize the benefits (from RQ1 and RQ2) of DP?}	
\end{itemize}

Our study \remove{is grounded in the expertise of two of our study team members in the foundations of DP. To supplement this knowledge, we conducted } consisted of semi-structured interviews with 19 data depositors, administrators, and analysts as they used a technical probe~\cite{hutchinson2003technology} -- a software prototype called DP Creator -- to make statistical releases, in order to understand perceptions, challenges, and opportunities around differentially private data analysis. \revision{The findings from these interviews were supplemented and framed by our study team members' expertise in the foundations of DP.}
Based on these sources, our study presents the following contributions:
\begin{itemize}
    \item Through interviews with these participants, we provide insight into challenges of applying DP to share or analyze datasets, such as: (1) understanding the reasoning behind and implications of decisions regarding privacy and utility, (2) conducting analysis without access to raw data, (3) assuming new risks and responsibilities, and (4) integrating DP with upstream and downstream data analysis pipelines. (Section~\ref{sec:challenges-using-dp})
    \item Our interviews also highlight the potential utility of DP for advancing goals of safe, open access to sensitive datasets. Participants were optimistic about expanding access to the general public, but pointed out challenges for using DP for research purposes. In particular, they expressed that training and expertise is still required to guide decision-making, and the constraints of DP make it hard to use for exploratory data analysis and replication of studies. (Section~\ref{sec:utility-of-dp})
    \item We discuss how DP requires modifications at every stage of the data science workflow. We discuss ethical, epistemic, and governance questions raised by DP for data science as a mode of knowledge production. We provide suggestions for integrating DP and data science, including: (1) more information flow between depositors and analysts, (2) consultation and guidance from experts in DP, (3) context-specific education, (4) governance of privacy-loss parameters through trained administrators. (Sections~\ref{sec:data-science-workflows}-\ref{sec:suggestions-reconfiguring})
    \item \revision{Finally, although not directly validated by our study design, we offer suggestions for research and design of DP data analysis tools, including: (1) providing more explanation behind selecting parameters, (2) creating workflows around trust and safety for both data practitioners and data subjects, and (3) developing features for automated and depositor-led data contextualization. (Section~\ref{sec:suggestions-design})}
\end{itemize}

The paper proceeds as follows. We begin in Section~\ref{sec:background} by explaining DP, its details in practice, and the DP Creator prototype. In Section~\ref{sec:related-work}, we discuss related work on this topic. Next, in Section~\ref{sec:methods}, we provide a description of methods, data collection, and analysis. In Section~\ref{sec:findings}, we discuss themes from our interviews and contextual inquiry. Based on these findings, we conclude in Section~\ref{sec:discussion} with a discussion about the ways in which the practices of data science itself must be reconfigured in order to comply with the constraints of DP, suggestions for more smoothly integrating DP and data science, \revision{and suggestions for design of DP data analysis tools}.


\section{Background}
\label{sec:background}

In this section, we provide background on differential privacy and the DP Creator prototype that we use as a technical probe~\cite{hutchinson2003technology} in our study. Based on our understanding of the foundations of DP, we also describe the information that interfaces for DP data analysis, such as DP Creator, require from data practitioners. 

\subsection{Differential Privacy}
\label{sec:dp}

Differential privacy (DP), introduced by Dwork, McSherry, Nissim, and Smith in 2006~\cite{dwork2006calibrating}, is a mathematical definition of privacy that limits how much information a mechanism for making a statistical release  reveals about any one individual in the dataset. 
The definition is parametrized by two quantities --- $\varepsilon$ and $\delta$ --- that denote the `privacy loss’ incurred by running a given set of analyses on the data.
In order to satisfy a guarantee of small privacy loss, the mechanism must introduce carefully calibrated noise to any computation over the data.

We provide the formal definition of DP below. Let $\mathcal{D}$ be a data universe and $\mathcal{D}^n$ be the space of datasets of size $n$. Two datasets $d, d' \in \mathcal{D}^n$ are neighboring, denoted $d \sim d'$, if they differ on a single record. Let $\mathcal{H}$ be a hyperparameter space and $\mathcal{Y}$ be an output space.	

\begin{definition}[\cite{dwork2006calibrating}]
A randomized mechanism $\mathcal{M}: \mathcal{D}^n \times \mathbb{R}_{\geq 0} \times [0,1] \times \mathcal{H} \rightarrow \mathcal{Y}$ is \emph{$(\varepsilon, \delta)$-differentially private} if for all datasets $d \sim d' \in \mathcal{D}^n$, privacy loss parameters $\varepsilon \geq 0$, and $\delta \in [0,1]$, hyperparameters $hp \in \mathcal{H}$, and events $E \subseteq \mathcal{Y}$,
\[
 \Pr[ \mathcal{M}(d, \varepsilon, \delta, hp) \in E] \leq e^{\varepsilon} \cdot \Pr[ \mathcal{M}(d', \varepsilon, \delta, hp) \in E] + \delta
\]
where all probabilities are taken over the random coins of $\mathcal{M}$.
\end{definition} 

The mathematical formalization of DP accounts for current and future attacks, remains robust to arbitrary auxiliary information, and measures compositions of privacy loss over multiple data releases~\cite{dwork2014algorithmic}. In addition, unlike heuristic approaches to privacy that rely on security by obscurity, DP enables third-party scrutiny of the algorithm, which makes it possible for a data analyst to take into account the noise introduced when performing inference and estimating uncertainty through confidence intervals.

DP has become a gold standard for measuring and controlling privacy loss for statistical releases on sensitive datasets. Disclosure methods that satisfy DP have been adopted by a variety of data-collection agencies and institutions, including Google~\cite{erlingsson2014rappor}, Apple~\cite{greenberg2016apple}, Uber~\cite{near2018differential}, and the U.S. Census Bureau~\cite{machanavajjhala2008privacy,abowd2018us}.

\subsection{Applying DP in practice}
\label{sec:dp-in-practice}
Using DP requires data practitioners \revision{(ie. data depositors and/or data analysts, as defined in Section~\ref{sec:dp-creator})} to provide information about the dataset or data domain, and to make choices about privacy and utility. Many conversations around using DP in practice focus on selecting the privacy-loss parameters ($\varepsilon$ and $\delta$), but in reality, setting these parameters is just one of several decisions and choices that practitioners must make during the data analysis process. Based on our collective expertise\revision{\footnote{\revision{We rely on this expertise to frame our main findings in this paper (ie. Sections~\ref{sec:findings}-\ref{sec:discussion}), which are primarily derived from our interviews with data practitioners.}}} conducting research in the foundations of DP, designing DP tools, and working with data practitioners, we made a list of \remove{other }parameters that practitioners may need to choose within a DP data analysis interface. These include the following:
\begin{itemize}
    \item \emph{Validating that the dataset is appropriate for DP}, for example, by confirming that it contains moderately\footnote{Data that is not very sensitive can potentially be made available without using DP, and data that is highly sensitive may need access control mechanisms beyond just using DP and additional review.} sensitive data about individuals.
    \item \emph{Providing information about how the data was sampled}, such as whether it was a simple random sample from a larger population (which, when supplied with the size of this larger population, would amplify the privacy guarantee~\cite{balle2018privacy}), or whether a data-dependent sampling scheme was used (which might degrade the privacy guarantee~\cite{bun2022controlling}).
    \item \emph{Selecting privacy-loss parameters}, $\varepsilon$ and $\delta$; the smaller the parameters, the more privacy is retained in the release. This is also called a privacy-loss budget.  
    \item 
    \emph{Selecting metadata parameters}, such as 
     ranges for numerical variables and categories for categorical variables. These parameters are necessary for limiting the amount of noise added to a statistic. It is important to set these inputs carefully for both \DP{utility and privacy}: overly large ranges or extensive categories may lead to high variance in the noisy statistic, while overly narrow ranges or limited categories may lead to biased outputs. Crucially, ranges and categories should \emph{not} be derived from the values in the dataset itself; data-dependent parameter selection constitutes a potential privacy risk. Rather, they should be set according to a \DP{codebook} for the dataset or \DP{knowledge about the population independent from the dataset.} 
    For example, the range for an ‘age’ variable could be set to 0-110 based on general knowledge about human lifespans, but it should not be set to values informed by the private dataset itself, such as 5-96.
    \item \emph{Allocating the privacy-loss budget amongst different statistics.} Researchers typically run many analyses on a single dataset. Using the composition theorems in DP~\cite{dwork2006calibrating,kairouz2015composition,murtagh2016complexity}, a software tool can \DP{analyze the privacy loss over multiple statistical releases.} However, the data depositor may wish to distribute a global \DP{privacy-loss budget} over many data analysts or releases, and the data analysts will need to distribute their allocated portion of the budget over multiple statistics.
\end{itemize}
The need to make these choices is well-known to experts in DP, but often gets lost in the discussions around deployments. As indicated by our findings in Section~\ref{sec:findings} and the discussion in Section~\ref{sec:discussion}, these decisions pose challenges for non-experts in DP throughout the data analysis process.

\begin{figure}[h!]
\begin{subfigure}{0.86\columnwidth}
  \centering
\includegraphics[width=\linewidth]{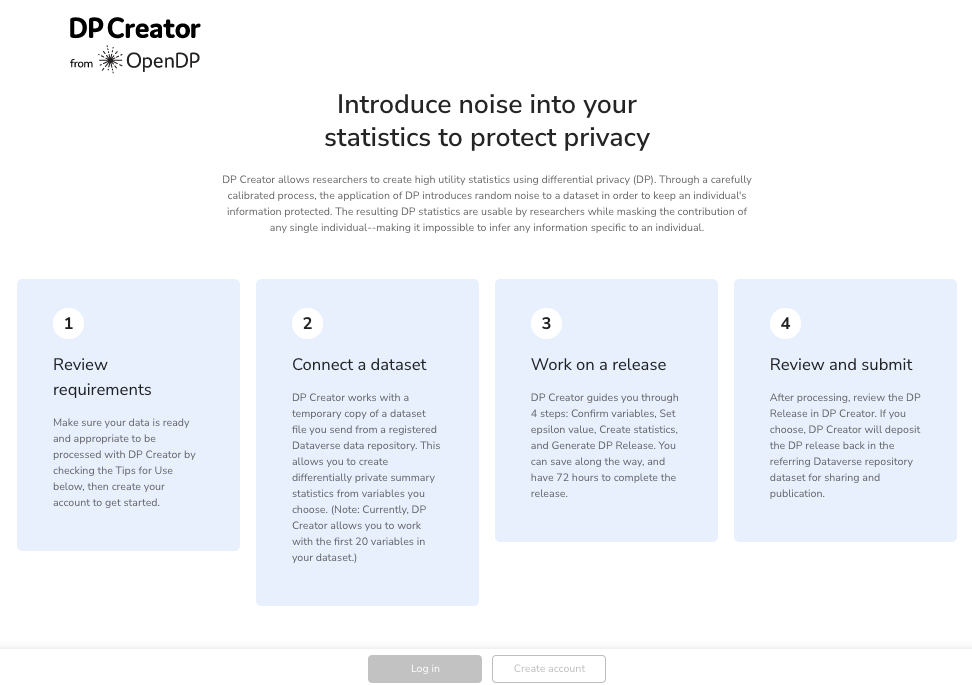}
  \caption{Overview of workflow}
  \label{fig:dp-creator-overview}
\end{subfigure}%
\hspace{0.1em}
\begin{subfigure}{0.86\columnwidth}
  \centering
\includegraphics[width=\linewidth]{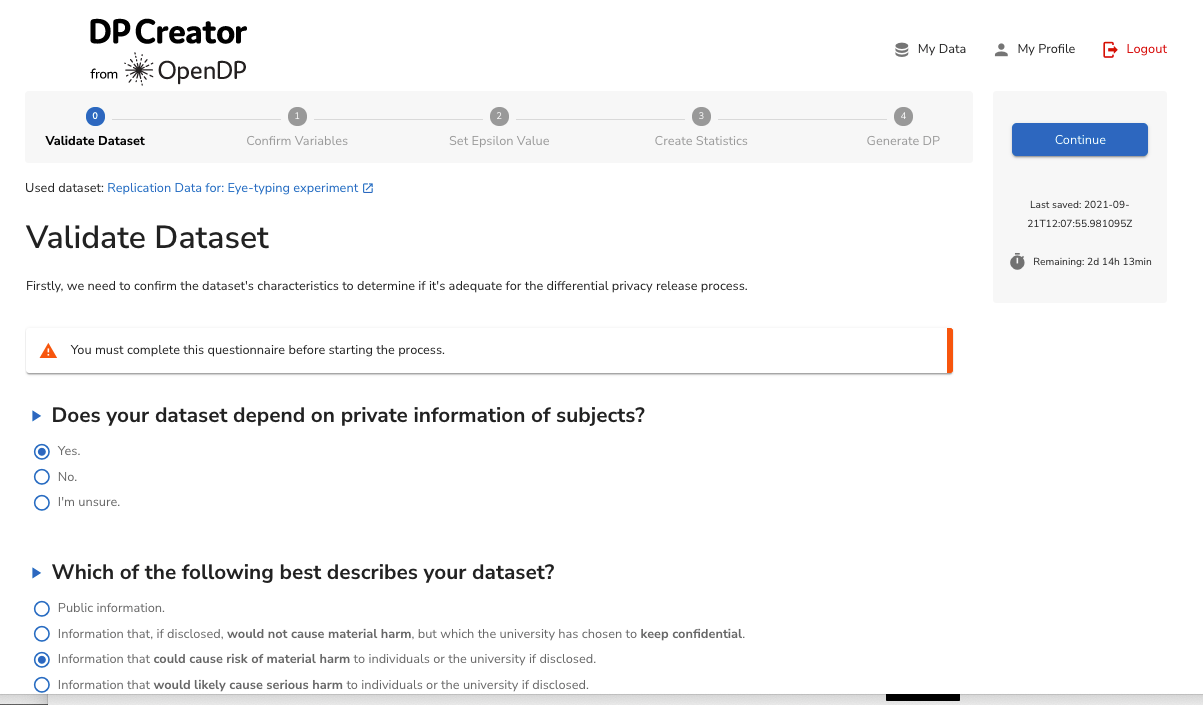}
  \caption{Validating dataset}
  \label{fig:dp-creator-validate}
\end{subfigure}
\begin{subfigure}{0.86\columnwidth}
  \centering
\includegraphics[width=\linewidth]{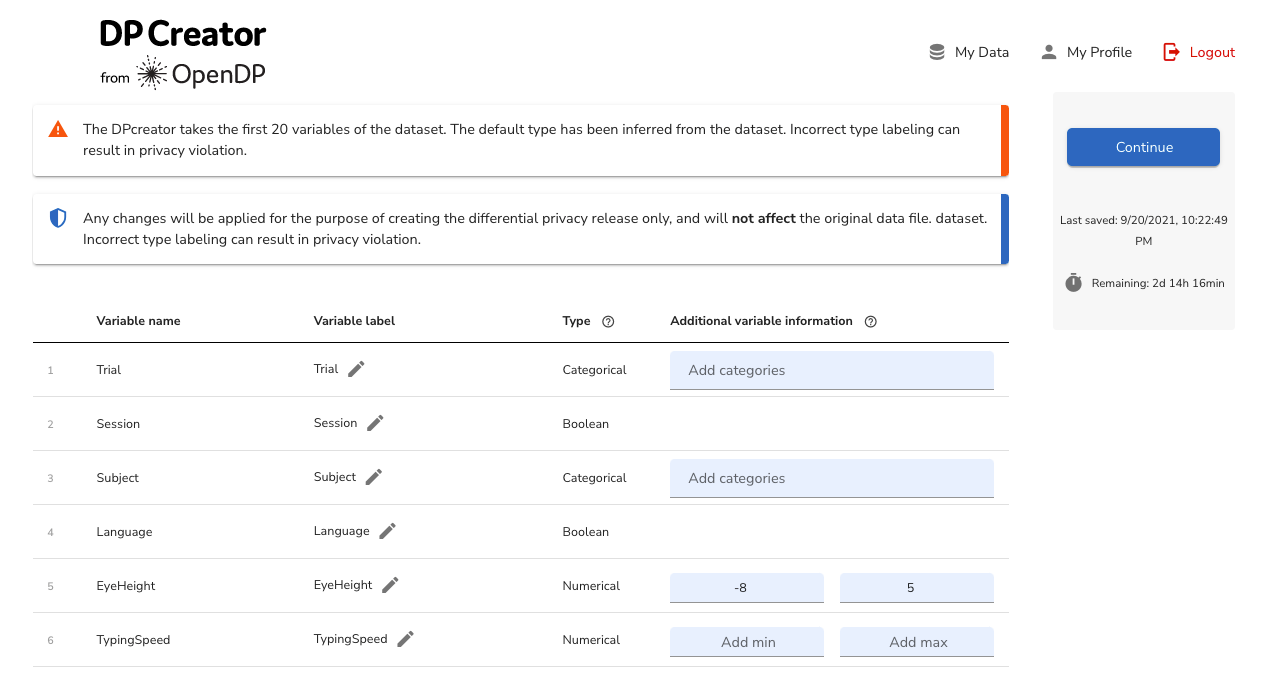}
  \caption{Selecting metadata parameters}
  \label{fig:dp-creator-metadata}
\end{subfigure}
\begin{subfigure}{0.86\columnwidth}
  \centering
\includegraphics[width=\linewidth]{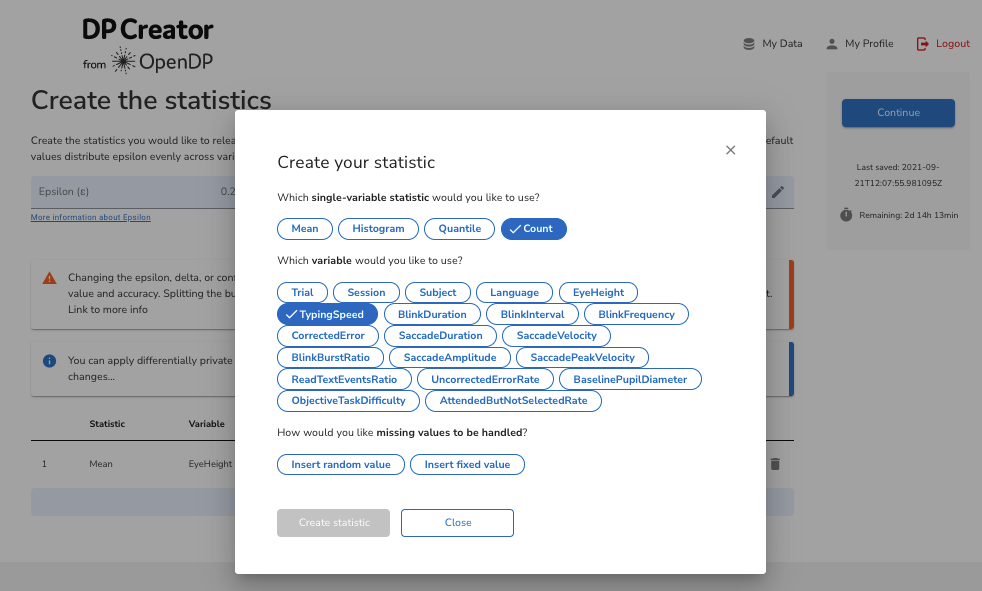}
  \caption{Creating a DP statistic}
  \label{fig:dp-creator-statistic}
\end{subfigure}
\caption{Some screenshots of the DP Creator prototype's data depositor interface.}
\label{fig:dp-creator}
\Description{Screenshots of the DP Creator prototype's data depositor interface.}
\end{figure}

\subsection{DP Creator}
\label{sec:dp-creator}

Our study involves observing participants as they use a prototype of a differentially private data analysis tool called DP Creator, which is part of the OpenDP open-source software project~\cite{gaboardi2020programming}. DP Creator is a software tool that aims to make it easier for people who are not experts in privacy, statistics, or computer science to reap the benefits of DP when sharing their data.
(Note that our study team is affiliated with OpenDP, and some of the authors were involved in designing the DP Creator tool.)
While other DP systems are designed to optimize utility for specific data releases or specific algorithms, DP Creator seeks to be a general-purpose tool for privacy-preserving data science that is compatible with existing workflows in social science research.\footnote{DP Creator evolved from a tool called PSI: a Private data Sharing Interface, the white paper for which~\cite{gaboardi2016psi} provides more details on the motivations and design of DP Creator. See also Murtagh et al.~\cite{murtagh2018usable} for an initial usability study of PSI.}
The tool has three interfaces according to three different user roles: data depositor, data analyst, and data archive administrator.

\begin{itemize}
    \item The \emph{data depositor} interface is for a researcher who has collected sensitive data about individuals and who would like to \revision{analyze and} share summary statistics about this data using the tool. \revision{The depositor also provides metadata information to facilitate future DP analyses of the dataset beyond her initial release.} \\
 
     \vspace{-1em}
 The data depositor uploads her data to the tool directly or through an online data repository such as Dataverse~\cite{magazine2011dataverse}. If possible, she should also upload a codebook or schema for the dataset. She must first validate that the dataset is appropriate for applying DP (as shown in Figure~\ref{fig:dp-creator-validate}): for example, it should contain moderately sensitive information about individuals, and the dataset should be organized such that each row corresponds to an individual (because DP provides an individual-level privacy-loss guarantee). Then, she must answer whether the dataset is a simple random sample from a larger population, and if so, input the size of the larger population.\footnote{The DP Creator prototype only asks users about simple random sampling; it has not yet incorporated recent work on data-dependent sampling schemes and their effects on the privacy guarantee~\cite{bun2022controlling}.} 
    Next, she must set the \revision{global} privacy-loss parameters for the dataset. 
    The prototype currently allows the depositor to select a ``data security classification" level for the dataset according to the organization's research data guidelines,\footnote{For example, see https://security.harvard.edu/data-security-levels-research-data-examples} which it then matches to pre-set values of $\varepsilon$ and $\delta$.
    \footnote{These parameters were decided and hard-coded by the designers of DP Creator. We find these parameters reasonable, but it is important to note that there is currently no consensus around how to select $\varepsilon$ and $\delta$ parameters, nor is it a widespread practice to set these parameters according to data security classification levels. For example, if the depositor selects ``Information that could cause risk of material harm to individuals or the university if disclosed," the prototype sets $\varepsilon = 1.0$ and $\delta = 10^{-6}$. For an in-depth discussion around selecting privacy-loss parameters, see Dwork et al.~\cite{dwork2019differential}. } 
    The depositor can also set or adjust these parameters manually.\revision{\footnote{\revision{The prototype offers guidelines for setting parameters manually, such as choosing $\varepsilon$ to be between $0.1$ and $1.0$ and $\delta$ to be smaller than $10^{-6}$.}}}  \revision{The depositor is prompted to leave a portion of the global privacy-loss budget for future analysts who would like to submit queries to this dataset.}
    Next, the tool will present the variables in the dataset, their labels and types, and any metadata information (such as ranges, categories, or descriptions) available in the codebook (See Figure~\ref{fig:dp-creator-metadata}). 
    The data depositor must adjust these ranges and categories according to her preferences and guidelines from the prototype; \revision{for example; the prototype explains that ranges that are too large will increase variance of the noisy statistic, while ranges that are too small will lead to biased outputs.\footnote{\revision{Note that the prototype does not offer a way for the depositor to select metadata parameters based on a well-defined calculation of the bias-variance tradeoff, but simply states that a tradeoff of this form exists.}} The prototype also reminds the depositor that these metadata parameters should not be selected based on information from the dataset itself.} Next, the depositor chooses which statistics to compute over subsets of the variables (See Figure~\ref{fig:dp-creator-statistic}). Once the depositor has chosen all of her desired statistics, she should adjust how much of the privacy-loss budget \revision{for this initial release} is allocated to each statistic based on how this affects the error on each statistic (spending more budget leads to less error on that statistic). Finally, the depositor confirms that she is ready to submit her statistics and is able to share the output of the DP analysis.
    
    \item The \emph{data analyst} interface is for a researcher or member of the public who does not have access to the raw data, but who would like to receive summary statistics about the data (perhaps to decide whether or not to go through the lengthy process of applying for access to the raw data). \\
    
    \vspace{-1em}
    First, the data analyst must set select the dataset she wishes to explore using DP. The prototype then presents the variables in the dataset, their descriptions, and any metadata parameters (such as ranges, categories, or descriptions) as inputted by the data depositor. The data analyst may also be able to refer to a codebook for the dataset, if this has been uploaded by the depositor.
    The data analyst must select the variables she wishes to work with, and can adjust the ranges and categories according to her preferences \revision{(again, the prototype explains to the analyst that a general bias-variance tradeoff exists with respect to the selection of these parameters).}
    The privacy-loss parameters \revision{allocated to this specific analysis} are already set by the data depositor; the data analyst cannot change these values. Then, the analyst chooses which statistics to compute over subsets of the variables. Once the analyst has chosen all of her desired statistics, she should adjust how much of the privacy-loss budget \revision{for this analysis} is allocated to each statistic based on how this affects the error on each statistic (again, spending more budget leads to less error on that statistic). Finally, the analyst confirms that she is ready to submit her statistics and receives the output of the DP analysis.
    \item The \emph{data administrator} interface is for a trained data curator or research data librarian who manages the different datasets and access policies for a given data archive. The data administrator, with consultation with data depositors, may be tasked with approving or denying requests from researchers to either access DP summary statistics released by the depositor, or to conduct their own DP analyses, on sensitive datasets. The administrator may also manage the allocation of the privacy-loss budget amongst different analysts. 
\end{itemize}
In this study, we consider the first two interfaces, as these are the ones available in the DP Creator prototype. We also consider the three main use cases that the prototype aims to support: exploratory data analysis, replication of scientific studies, and wider access of sensitive datasets for the public~\cite{gaboardi2016psi}. The current version of the prototype and a tutorial\footnote{The tutorial for using this prototype can be found at the following link: \url{https://docs.google.com/document/d/e/2PACX-1vRlZ2IgigIhl4oz_uOakQPxovzlrmFkbD-x_9RUO31dC0eRq2wCt_vN2Go0_9LTRd67srjgy04CfPVk/pub}} can be accessed at the following link: \url{https://demo-dataverse.dpcreator.org/}; note that this version may have minor changes compared to the version of the prototype used in our study.



\section{Related Work}
\label{sec:related-work}
We situate our work within three areas in privacy and HCI literature: privacy communication, defining users of privacy tools, and translating ethical data science from theory to practice.

\subsection{Privacy Communication} 
Several recent works have explored communication of DP guarantees to data subjects who may be faced with the decision of whether or not to have their data collected.
Bulleck et al.~\cite{bullek2017towards} explore how data subjects understand and interact with privacy-loss parameters for survey responses. After being given a sensitive question, participants were asked to choose a level of perturbation (ie. privacy) to add to their response. Interestingly, some participants chose the lowest level of privacy for their response, revealing the unexpected behavior that can emerge when participants do not understand the norms of practice that are implicit within a given privacy technique. 
Xiong et al.~\cite{xiong2020towards} explore data disclosure decisions from data subjects, elucidating reasons that users choose to share their data. They find that more explanation of the implications of DP yields further sharing of sensitive data; however, they also find that even when users claim to understand the explanations, objective measurements show that the rates of comprehension are quite low. 
Cummings et al.~\cite{cummings2021need} also examine the dynamics of comprehension and data sharing from data subjects. They find that users care about different kinds of information leakage, and that the manner in which DP is communicated sets privacy expectations accordingly. 
Most recently, Smart et al.~\cite{smart2022understanding} investigate explanations of DP that hide important information such as privacy-loss parameters. Interestingly, they find that explanations have little effect on individual's willingness to share, which is determined even before they learn about the privacy protections at hand.

These works offer important lessons for communication of DP guarantees to end users. They point out several ways in which users can misunderstand the common explanations used regarding DP, and they caution that we must be intentional with the way explanations are conveyed, for such explanations can affect participants' motivations for data sharing in subtle or unexpected ways. However, while these works consider communication for end users who are data subjects, our work takes end users to be DP practitioners who may not have the ability to ``opt out" if they are unsure about the protections offered, as data subjects often can, but rather must make critical decisions about how to handle people's sensitive data.


\subsection{Defining users of privacy tools} 
Although less studied, recent work has started to examine other perspectives of users who are not data subjects.
Agrawal et al.~\cite{agrawal2021exploring} argue that when considering privacy-preserving computation, which is typically far removed from the data subject or app user, the notion of ``user” must be reconceptualized to include designers, developers, and policymakers who work closely with privacy-preserving computational tools. 
Based on interviews with these parties, 
the authors find that tools for privacy-preserving computation remain a mystery to many who interact with them. This poses pressing questions for governance and design of these tools. Our work argues that in the case of DP, the tool may similarly produce unexpected behavior from \remove{users}\revision{data practitioners} (that can be detrimental to privacy and/or utility) unless norms of practice are conveyed both within and outside of the user interface, through direct engagement and context-aware educational guidance. 

Nanayakkara et al. ~\cite{nanayakkara2022visualizing} also consider the perspective of research practitioners in their study. They interview participants and evaluate their comprehension as they use a visualization tool to explore the relationships between $\varepsilon$, accuracy, and disclosure risk, as well as the impact of DP noise on statistical inference. This work underscores the importance of tools for practitioners to understand the dynamics between privacy and accuracy and the implications of making parameter choices in a hands-on manner. Their tool is complementary to the prototype we use for this study, and our findings show that participants would benefit from being able to visualize these tradeoffs. However, as the authors point out, more work needs to be done to understand how generating visualizations cuts into the privacy-loss budget, and how to use the visualization to actually perform a DP analysis.

Concurrent work by Munilla Garrido et al.~\cite{garrido2022lessons} investigates the `academic-industrial DP utilization gap' through interviews with data analysts and data stewards in major companies that have not yet deployed DP. Their findings about the barriers to adoption of DP, as well as the promises of using DP, support the findings in this paper. In particular, Munilla Garrido et al. conclude that DP can simplify onerous data access processes and facilitate data exploration across silos. They are optimistic about bridging the divide between theory and practice of DP, identifying key technical gaps that DP tools should address in order to promote adoption across industry. Echoing findings from a previous interview study by Dwork et al.~\cite{dwork2019differential}, Munilla Garrido et al. emphasize the need for sharing learnings across the DP community.

Finally, Qin et al.~\cite{qin2019usability}'s work on multi-party computation (MPC), while not about DP, offers valuable insights on how to educate stakeholders about technical privacy guarantees ``while [keeping practitioners] insulated from the nuances of [the protocols'] implementation.'' Qin et al. find that describing simple examples of the MPC protocol to practitioners allows them to trust the protocol's security on more complicated functions. In addition, modeling the user interface on something that is familiar, such as a spreadsheet, allows practitioners to participate more confidently. Finally, Quin et al. emphasize the importance of collaboration between security engineers and human factors experts in designing interfaces, particularly in settings (including MPC and DP) where users can only interact with the interface only a limited number of times.

More generally, scholars have highlighted the ways in which computational tools shape the identity of users as well as the scope of their actions. Grint and Woolgard~\cite{grint1997machine} describe this process in detail, highlighting that ``by setting parameters for the user’s actions, the evolving machine effectively attempts to configure the user.” We build on these insights in this work to show how DP tools may seek, but fail, to reconfigure data analysts in the desired manner. We argue that reconfiguring users to make safe, ethical choices in the data analysis process requires not only better design within the tool, but also robust guidance, training, and education outside of the tool.


\subsection{Ethical data science} 
Finally, there has been a slew of recent work on the challenges of translating ethical data science from theory to practice. Seda Gürses~\cite{gurses2016privacy} draws attention to the “scant attention given to socialization of [privacy] tools” as an obstacle for bringing privacy-preserving methods from the lab into the world. Jörg Dreschler~\cite{drechsler2021differential} highlights several ways in which differentially private data analysis clashes with fundamental workflows and modes of interaction within government statistical agencies. Oberski and Kreuter~\cite{oberski2020differential} similarly delineate the tensions between DP and social science, such as conceptions of risk and workflow, but they also outline ways in which DP may lead to better social science by protecting against p-hacking and introducing robustness into the data analysis process. Barocas and boyd emphasize the tensions between ethicists who critique the ‘ethically uninformed’ practices of data science, and data scientists who feel that they already embed nuanced ethical considerations into every step of their analyses~\cite{barocas2017engaging}. Finally, boyd and Sarathy~\cite{boyd2022differential} describe the challenges faced by the Census Bureau in communicating about DP as it conflicts with dominant `statistical imaginaries' about census data.

These works show that there are significant gaps between theory and practice of ethical data science which are crucial to explore further. Our work offers insights into barriers to adoption of differentially private data science, and discusses how we can pave the way towards a smoother uptake of DP in practice.

\section{Methods}
\label{sec:methods}

Our research approach consisted of both a \remove{contextual inquiry}\revision{think-aloud protocol} as well as in-depth, semi-structured interviews with data practitioners. As DP is still considered an emerging technology, 
and there has not been much previous literature that empirically considers DP tools from the perspective of data practitioners,
our research questions were open-ended and exploratory. At the same time, since our goal was to explore DP in practice, we decided to engage our participants using a technology probe~\cite{hutchinson2003technology}. We observe participants as they use DP Creator, the prototype of a differentially private data analysis tool described in Section~\ref{sec:dp-creator}.

\begin{table}[]
\centering
\begin{tabular}{|c | c | c |} 
 \hline
 \textbf{Participant} & \textbf{Role} & \textbf{Background} \\ [0.5ex] 
 \hline
 D1 & Depositor & Researcher \\ 
 \hline
 A2 & Analyst & Researcher \\
 \hline
 A3 & Analyst & Researcher \\
 \hline
 D4 & Depositor & Researcher  \\ 
 \hline
 D5 & Depositor & Researcher \\ 
 \hline
 A6 & Analyst & Researcher \\ 
 \hline
 D7 & Depositor & Researcher \\ 
 \hline
 A8 & Analyst & Researcher \\
 \hline
 A9 & Analyst & Researcher \\ 
 \hline
 A10 & Analyst & Data archive administrator\\ 
 \hline
 D11 & Depositor & Data archive administrator \\ 
 \hline
 D12 & Depositor & Data archive administrator \\
 \hline
 D13 & Depositor & Data archive administrator \\ 
 \hline
 A14 & Analyst & Data archive administrator \\ 
 \hline
 A15 & Analyst & Data archive administrator \\ 
 \hline
 A16 & Analyst & Data archive administrator \\ 
 \hline
  D17 & Depositor & Data archive administrator \\ 
 \hline
  D18 & Depositor & Data archive administrator \\ 
 \hline
 A19 & Analyst & Data archive administrator \\ 
 \hline
\end{tabular}
    \caption{Summary of total participant sample.}
    \label{tab:participants}
\end{table}

\subsection{Recruitment}
\label{sec:recruitment}
Participants were recruited through a combination of convenience and purposive sampling~\cite{etikan2016comparison}. We sought participants who were not experts in DP, but who were experienced in working with sensitive data. This included researchers from labs that had attempted to expand access to their own sensitive datasets, researchers who had experience applying for access to and conducting analyses on sensitive data, and data archive administrators who managed access to sensitive datasets. We relied on the network of the open-source software project OpenDP~\cite{gaboardi2020programming} \remove{which is well-connected with organizations and individuals whose use-cases may be suitable for differentially private data analysis, in order to identify potential participants.} \revision{in order to identify organizations with use-cases of data that may be suitable for differentially private data analysis. In these organizations, there was typically one or two contacts who were well aware of and at least somewhat experienced with differential privacy. We asked these contacts to point us to other members of the organization who had little to no prior experience with differential privacy. (See Section~\ref{sec:screening-questions} for an example of screening questions.) For example, one research group leader we reached out to through the OpenDP network responded to us that ``I think [name of analyst in their research group]  will be perfect - very data savvy, but without direct experience with DP.''}
\revision{Then, through open-ended exchanges with these potential participants, we asked them to tell us about their familiarity with differential privacy in order to confirm, using our own judgment, that the participants had no direct experience with or training in DP; for example, as one of our participants said, they ``have heard of DP but don't know what it means or how it works." }


We interviewed 19 participants in total: 9 academic researchers in social science disciplines
and 10 curators of research data repositories across a few different US and international institutions (for example, Dataverse repositories~\cite{magazine2011dataverse}). 
The summary of the total participant sample is shown in Table~\ref{tab:participants}.

\revision{
\subsection{Data}
\label{sec:data}
We asked participants who were researchers (9 out of 19) to provide simulated or actual sensitive data that their research group had collected\revision{; out of these, 2 provided real data, 4 provided simulated data, and 3 did not provide data.} 
\revision{The 2 participants that supplied real data were part of the same research group, which had collected this data as part of an online service deployed by the group.} 
\revision{See Section~\ref{sec:data-ethics} for more details on ethics considerations around the use of these data and measures taken to protect data subjects.}
\revision{The 4 participants that supplied simulated data were all part of a second, shared research group, and this synthetic data was created (using a statistical model applied to the real data) to have similar characteristics to their groups' real data.}
For the rest of the participants (3 researchers and 10 administrators), we uploaded into the prototype a 2010 American Community Survey (ACS) Public Use Microdata Sample (PUMS) dataset that contains demographic and income information about California residents.\footnote{Such samples are publicly available at \url{https://www.census.gov/programs-surveys/acs/microdata.html}. Note that ACS PUMS files are already privacy-protected using disclosure avoidance techniques applied by the Census Bureau. However, for the purposes of our study, we treated this dataset as containing unprotected, sensitive information about individuals. In addition, note that ACS data are not weighted according to a simple random sampling design, but this was not taken into account within the prototype. To learn more about ACS design and methodology, see \url{https://www.census.gov/programs-surveys/acs/methodology/design-and-methodology.html} }
}

\revision{We ensured that all of the datasets were suitable for use with differential privacy in terms of size, sensitivity, and format. The sizes of the datasets varied, but they were all large enough to provide good utility for differentially private statistics ($> 5,000$ observations) yet small enough for the DP Creator prototype to handle efficiently ($< 50,000$); we asked the dataset holders to provide a subset of the data if it was greater than this size. The datasets all contained demographic information about (possibly simulated) individuals that could be potentially sensitive such as age and sex, and therefore required some privacy protection, but they did not contain attributes that would be considered highly sensitive such as medical information. Finally, all of the datasets were formatted such that each row corresponded to one individual and all of their attributes; this allowed for a standard application of differential privacy without additional dataset transformations.
}

\subsection{Study Protocol}
\label{sec:study-protocol}
\revision{Our study protocol varied based on each participants' background, role, and data used. Below, we describe our general study protocol and where we made these variations.}

We began by asking participants about their skills, background and experience. Based on this information, we assigned her a role as a data depositor or data analyst. For example, if a researcher had spent significant time working with the particular dataset provided by their group for the study, we assigned her to the role of data depositor. Otherwise, if the researcher was not very familiar with the dataset, we assigned her to the role of data analyst.  These role assignments mirrored a real-life scenario, in which a data depositor would be very familiar with the dataset, while an analyst would not have had any prior access to the dataset but would be familiar with the data domain.\footnote{For example, the data analyst might be familiar working with ACS PUMS files regarding New York, but not California. Therefore, she might be familiar with how the data is coded and collected, but not with the relationships present in the data that may be specific to California.} \revision{As our screening questions (see Appendix~\ref{sec:screening-questions}) specifically asked contacts to point us to members of their research groups with high and low experience with the group's dataset, we were able to categorize researchers into these the roles of depositor and analyst without much ambiguity.\footnote{\revision{In smaller organizations, individuals may carry out the roles of both data depositor and data analysts, so there might not be much distinction between these roles. For the participants in our study, however, we found that there was only a one-directional overlap: the depositors nearly always participated in analysis, but the analysts were not always depositors. This allowed us to categorize participants accordingly.}}} For the 3 remaining researchers and 10 data archive administrators, we assigned the role of depositor or analyst based on which of the two roles the participant had more experience working in or with; if this was unclear, we made chose a role that would balance out our numbers of depositors and analysts. Our total sample contained 9 depositors and 10 analysts.

Next, the participant was given a role-dependent task to carry out using DP Creator. If she was playing the role of a data depositor, we asked her to use DP Creator to release useful insights about the sensitive dataset to the public. If she was in the role of a data analyst, we asked her to use DP Creator to explore and discover new insights about the data. These tasks were designed to give the participant freedom while using the tool and to match the free-form environment that the participants are accustomed to when working with data~\cite{tukey1977exploratory,dasu2003exploratory}. If the participants' research group had provided data for us to use, we knew that those participants were either familiar with the dataset itself and/or the data domain, and so we concluded that they would have a decent understanding as to which statistics they would want to release or explore. For participants using the California PUMS dataset, we gave them the following task (wording tailored for depositors):

\begin{quote}
Imagine that you are a researcher who has collected sensitive information about individuals in California, including attributes such as sex, race, marital status, and income level. Your dataset contains a simple random sample of 30,000 individuals from a population of 30 million individuals in California. Your goal is to use the DP Creator prototype to release privacy-preserving summary statistics that convey the main insights of your data. We’d like you to use the prototype to generate the following differentially private statistics: 
\begin{enumerate}
\item Mean age of individuals in California; missing values should be replaced with the number 30.
\item OLS regression using $x$=age and $y$=sex as variables; missing values for age should be replaced with 30 and missing values for sex should be dropped.
\end{enumerate}
\end{quote}

We used a think-aloud protocol~\cite{lewis1993task} to observe participants as they were interacting with the prototype: we encouraged participants to vocalize their perceptions, questions, and challenges while using the prototype. 
Once participants completed the task, we asked them a series of semi-structured, open-ended reflection questions, which were adapted based on the particular session and participant. The questions were geared towards understanding participants' thoughts on using DP to share and analyze sensitive data, the utility of the DP releases, and potential use-cases of DP for providing open access to datasets in a safe manner. (See Appendix~\ref{sec:sample-questions} for sample interview questions.) The \revision{think-aloud protocol} combined with open-ended interview  allowed us to explore both specific and high-level insights, many of which were relevant beyond the DP Creator prototype.

Interviews were recorded and transcribed; 
they lasted between 45 and 120 minutes, producing about 20 hours of audio and video recordings. 
The study team also took notes during the sessions. All parts are approved by an IRB.

\revision{
\subsection{Ethical Considerations} 
\label{sec:data-ethics}
Two out of the 19 participants in our study used real data from their shared research group as input to the DP Creator prototype. We obtained permission from the research group leaders and participants to use this data and made sure that the consent form for the data collection permitted the use of the data in our study. Most importantly, we carefully considered the ethics of using this data from the point of view of protecting data subjects in our study. Below, we describe below the measures taken towards this goal.}

\revision{Both participants used the same  dataset, which was collected as part of an online service that their research team had deployed. 
The terms of service and privacy policy\footnote{\revision{In order to maintain the anonymity of our participants, we are not able to attach these terms of service as supplementary material. However, we describe the relevant portions here.}} allowed for the sharing of personally identifiable information with other researchers for broad purposes, including improving the service for which the data was collected and for scientific research. The terms also allowed for the sharing of aggregate statistics that are not personally identifying with third parties and the public. However, in considering the ethics of using this real data for our study, we took steps well beyond what the terms of service and privacy policy required.}

\revision{
In particular, we made sure that the sensitive data was never exposed in the clear to anyone outside of the aforementioned two participants who brought this data to the study session. First, the DP prototype was installed on each of the participant's machines and the analyses were conducted locally, which ensured the dataset remained in sole possession of the participants. This avoided any cloud or network security concerns with storing information about the dataset. Second, our study team never viewed, had access to, or collected any information directly about the sensitive data. We only saw or recorded what was visible through the DP Creator interface, which included the size of the dataset, metadata parameters including the variable names and types, and the differentially private outputs which we discuss more below.}

\revision{
We also made sure to limit anything the study team may have been able to glean about the sensitive data \emph{indirectly}, which could have come from two sources: (1) the participant's actions and think-aloud protocol, and (2) the differentially private aggregate statistics about the data that were produced by the DP Creator prototype. For (1), we reminded participants before starting that they should not discuss anything about the sensitive dataset that would reveal personally identifying information and/or information that was not already publicly available about the dataset. For example, the participant could talk to us about the accuracy of the differentially private analysis in qualitative terms, but they should not precisely indicate its relationship to the exact, non-DP statistic. If they had inadvertently done so, this information would have been removed from the recording, notes, and transcripts; however, this did not happen. For (2), we made sure that participants only used conservative privacy-loss parameters ($\varepsilon = 0.1, \delta = 10^{-6}$) before proceeding with the analysis, and as the participant was moving through the workflow of the prototype, we corrected any errors they had made that could compromise the privacy guarantee in order to ensure that the differentially private aggregate statistics robustly preserved the privacy of data subjects. Out of an abundance of caution, we also made sure that our recordings, notes, and transcripts did not include the value of this final output, even though it was already protected by DP.}

\revision{Finally, for the simulated data provided by 4 participants from their shared research group, we took similar but slightly less stringent measures. This data was fully synthetic, and had already gone through robust disclosure avoidance procedures, but we still wanted to be careful in protecting the privacy of these (simulated) data subjects. Therefore, we again made sure to comply with the data use agreement, obtain permission from the research group, only view and record what was visible through the DP Creator interface, and limit what we may have learned about the dataset indirectly from our participants.}
\subsection{Data Analysis} 
The notes, transcripts, and audio/video recordings from the sessions were analyzed by two of the study team members using a reflexive thematic analysis approach adapted from Braun \& Clarke~\cite{braun2021one}. One of the two team members has expertise in the foundations of DP; the other was familiar with the broad concepts of DP, and had expertise in designing and evaluating user interfaces. The team members each spent extended time familiarizing themselves with the materials from each interview, 
taking down notes on additional points of interests to direct the analysis and reflecting on our own positionality with regard to DP data analysis. Guided by the research questions and these additional notes, the team members open-coded the data to develop semantic and latent sets of codes reported in Appendix~\ref{sec:codebook}.  Examples of codes from the interviews included: \textit{difficult to check assumptions about data}, \textit{worried about being held liable for exposing data}, and \textit{misconception about how metadata parameters affect privacy}. 
These two authors then clustered the codes into sets of themes for each research question.\footnote{As RQ3 considers broader implications of DP, our themes for this question integrated higher-level codes with our \remove{own expertise and }review of the literature.} Over multiple discussions and iteration with the broader study team, these themes were refined into the final set of themes summarized in Table~\ref{tab:findings} and reported in Section~\ref{sec:findings}. 


\section{Findings}
\label{sec:findings}

\begin{table*}[]
    \begin{tabular}{p{0.25\textwidth} p{0.4\textwidth} | p{0.35\textwidth}}
        \hline 
        \textbf{Research Question} & \textbf{Finding} & \textbf{Implications} \\
        \hline 
        \multirow{6}{12em}{What barriers do non-experts in DP face when using DP to share or analyze sensitive datasets? (RQ1)} 
        & Understanding the reasoning behind or implications of making choices (Sec~\ref{sec:understanding-complexities}) 
        & \revision{\multirow{8}{16em}{Design and research around DP data analysis tools (Sec~\ref{sec:suggestions-design}): \\ 
        (1) Provide more explanations behind selecting parameters and making decisions \\
        (2) Create workflows around trust and safety for both data practitioners and data subjects \\
        (3) Develop features for automated or depositor-led data contextualization  }} \\
        & Conducting analyses without access to raw data (Sec~\ref{sec:raw-data}) \\
        & Assuming new risks and responsibilities (Sec~\ref{sec:risks-responsibilities}) \\
        & Integrating DP into data analysis pipelines (Sec~\ref{sec:pipeline-integration}) \\
         \cline{1-2}
        \multirow{8}{12em}{
        \revision{What do data practitioners perceive to be the potential utility of DP for expanding access of sensitive data to the public, facilitating exploratory data analysis, and enabling replication?} (RQ2)} 
        & \\
        & Beneficial for wider access to the public (Sec~\ref{sec:wider-access})  \\
        & Poses challenges for exploratory analysis (Sec~\ref{sec:exploration}) \\
        & Will not necessarily enable better replication of scientific studies (Sec~\ref{sec:replication}) \\
        & Requires training, expertise, and governance (Sec~\ref{sec:expertise}) \\
        & & \revision{\multirow{8}{16em}{Differentially private data science (Sec~\ref{sec:suggestions-reconfiguring}): \\
         (1) Information flows from data depositors to analysts \\
         (2) Guidance from DP experts \\
         (3) Context-specific education \\ 
         (4) Governance of privacy-loss parameters }} \\
         & \\
         \cline{1-2} 
         \multirow{6}{13em}{ 
         \revision{What changes need to be made in the data science workflow to overcome the barriers from RQ1 and achieve benefits from RQ2?} (RQ3)}
         & DP changes every aspect of the data analysis workflow. (Sec~\ref{sec:data-science-workflows}) \\
         & Socializing users around the constraints of DP raises questions around ethical and epistemic implications 
         (Sec~\ref{sec:ethical-epistemic-implications}) \\
         &~\\
         \hline
    \end{tabular}
    \caption{Summary of research questions, main findings, \revision{and implications} regarding the use of DP by data practitioners for providing open access to sensitive datasets.}
    \label{tab:findings}
\end{table*}

\subsection{Challenges using DP}
\label{sec:challenges-using-dp}
In this section, we discuss the themes from interviews related to barriers that non-experts in DP face when using DP to share or analyze sensitive datasets (RQ1). We specifically highlight differences based on participants' roles (depositor or analyst). 
Overall, participants were optimistic about the possibilities afforded by DP to share data in currently-restricted settings while minimizing risk. However, they faced several challenges when using the tool, including not understanding the reasoning behind or implications of making certain decisions, working without access to raw data and only being allowed to perform computations through a black-boxed system, assuming new risks and responsibilities for protecting data subjects’ privacy, and not being able to integrate DP into their data analysis pipelines. 


\subsubsection{Participants grasped the broad concepts of DP, but struggled to understand the reasons or implications of their choices. }
\label{sec:understanding-complexities}

Participants came in with low experience with DP, but were able to understand and apply the concept of the privacy-loss budget when releasing statistics. In addition, participants seemed to grasp the \DP{relationship between the privacy-loss budget and the accuracy of statistics (ie. using a larger portion of the budget for any given statistic means greater accuracy for that statistic}, as demonstrated by their outward ruminations of whether or not adding an extra statistic would be worth thinning out the privacy-loss budget. 

However, participants were confused about why they needed to set some of the metadata parameters within the tool, even after reading the given explanations within the tool.
\begin{quote}
  Why are these parameters necessary? Where should I be getting this information? (A10)
 \end{quote}
 \begin{quote}
  Why do we need to set these parameters? I have no idea. Why am I telling you what you already know? (D11)
\end{quote}
A few participants seemed to conflate setting these metadata parameters with setting the privacy of different variables. Recall that the privacy loss of any release is only determined by the privacy-loss parameter and not by other parameters. Yet, even after setting privacy parameters, some participants still tried to influence the privacy loss of certain variables through decisions they made in the tool.
For example, while choosing the ranges for a variable of `income,' one of the participants, believing that income is a more sensitive variable than age, inputted a larger range as they believed this would lead to noisier statistics that would ``hide" the true values (A3). Another participant explained the mental gymnastics they were going through while making choices regarding privacy and accuracy:
\begin{quote}
    Someone might think, oh, this [variable] is sensitive, I need to hide it. The headspace is to leave out what is sensitive, but the \emph{goal} is to release sensitive information. So I need more guidance to clarify how to set these parameters. (D12)
\end{quote}
Within the prototype, depositors are asked to set privacy-loss parameters, while analysts are not. 
The amount of technical detail around these parameters that depositors could tolerate varied based on their background. One participant playing the role of depositor stated that as a social scientist with a limited mathematics background, she felt that the display of the $\varepsilon$ and $\delta$ privacy-loss parameters made her ``feel uncomfortable" (D1). Another participant stated that showing the technical meaning of these parameters was unnecessary: “Not being a specialist, I would go with the default values anyways” (D11). A third participant emphasized that given his less technical background, he had trouble understanding the meaning and implications of the privacy-loss parameters.
\begin{quote}
  $\varepsilon$ is easier to understand, but the $\delta$\ldots? Its abstract. Doesn’t mean much to me\ldots because I don’t study probability, this page would give me a lot of pause. (D13)
\end{quote}

On the other hand, participants who were better versed in probability and statistics were less fazed by seeing these parameters: “I have a rough understanding of $\varepsilon$ and $\delta$, but felt pretty good about it. Because I understood the general concept.” (D17) In fact, these participants expressed that they would have liked \emph{more} technical details about DP to be included in the tool. They felt that with the current amount of information presented, they did not have a strong enough conceptual understanding in DP to feel confident in using the tool in real life. 
\begin{quote}
    When choosing the privacy-loss budget, the guidelines were helpful, but I'd want more information for people like me who are kind of familiar with technical concepts like this. I'd also want more information about the utility-risk tradeoff. (D5)
\end{quote}
Even on the analyst side, participants well-versed in statistics asked for more information about the technical actions that users were implicitly taking within the prototype, such as running the noise-infusion algorithm.
\begin{quote}
    Not being fully trained with noisy data, I think it’s important to know how noise is added. (A6)
\end{quote}
Several of the participants also wanted more information about the implications of over or underestimating the parameters they were asked to input.
They suggested including references to educational resources, such as links to a DP textbook, peer-reviewed research papers, or a video overview of DP, for those who would be interested in learning more. We discuss educational resources in more detail in Section~\ref{sec:suggestions-reconfiguring}.


\subsubsection{Participants found it challenging to make decisions and interpret results without access to raw data.}
\label{sec:raw-data}

One of the major differences between DP tools such as DP Creator and non-private data analysis tools is that in DP Creator, data analysts do not have access to the raw data when exploring and accessing statistics. This poses a challenge to analysts who are accustomed to navigating the raw data to guide their analysis process. 

First of all, data analysts struggled with the data preparation and validation stage: not being able to directly clean the data, deal with missing values, and check their assumptions about the data.
\begin{quote}
    I'm used to seeing what the dataset looks like\ldots that can help you see how things like missing values should be treated. (A2)
\end{quote}
Analysts emphasized that running sanity-checks on the data is a key part of their analysis process. Not being able to do so in the same way that they were used to posed a real challenge.
\begin{quote}
    It's weird that I can’t see data, like I can’t check anything. Usually there’s always little things you do with the data to make sure things look right, such as tabulate the data to make sure missing values coded properly. It's weird to not have the distribution of values. (A8)
\end{quote}

Unlike data analysts, data depositors are permitted to see the raw data (as they have often collected it themselves). However, DP requires that any data-dependent decisions must be accounted for within the privacy-loss budget. Doing such accounting is difficult, however, and the prototype did not support this functionality. Therefore, we encouraged data depositors to simply not look at their own data throughout the entire process as a way of limiting their use of data-dependent knowledge. As discussed in Section~\ref{sec:dp-in-practice}, we instead encouraged depositors to use their knowledge of the data domain (e.g. that human ages are typically between 0 and 110) rather than dataset-specific information (e.g. exact min and max ages in the dataset) to set metadata parameters.

Information related to the data domain, rather than the specific dataset, can often be found within codebooks or schema. 
Some of our participants did bring and refer to a codebook of their group's dataset, but they found the codebook to often be inadequate in providing all of the desired context. When they were not permitted to look at the data at all, they needed much more annotation.

In the case where no codebook is available, the DP Creator prototype fills in some of the gaps; it currently compiles all of the names and brief descriptions of the variables in the dataset and displays this information to the user. However, especially for data analysts who do not have access to the raw data and are not familiar with the data at all, they experienced trouble interpreting the meanings of variables strictly from their names and descriptions:
\begin{quote}
    It would be helpful to be able to see the data variables' detailed descriptions here again. I assume this is annual income, but I’m not sure. Same with education. (A9)
\end{quote}
Even data depositors involved in collecting the data themselves struggled, for as one participant told us, "researchers don’t always remember, or there may be big teams [working with the data], realistically" (D13).
\begin{quote}
    If I was really going to do this, I need the variable description. Researchers can be bad at variable naming. This is a two-monitor job.  I would have to have another monitor open with my data, my R cheat sheet, and notes to update my codebook because I updated a variable. (D13)
\end{quote}

Not being able to look at the raw data while doing the analysis was especially challenging when participants were trying to input metadata parameters. Several depositors and analysts were unsure how to fill in the lower and upper bounds of numerical variables and resorted to educated guesses that ended up being far off from the actual bounds, causing the corresponding releases to be \DP{minimally informative.} For example, one of the analysts had been working closely with data similar to the sensitive dataset for a month prior to our interview, yet they could not figure out without looking at the dataset whether a key variable was coded from 0 to 1 or from 0 to 100. This made it difficult to input a reasonable range for this variable, which affected the utility of the outputs. At the end of the session, the analyst realized their mistake:
\begin{quote}
The histogram I got out at the end was not interpretable because I overshot the upper bound. (A8)
\end{quote}
If this had happened in a real-life scenario, the analyst would have to spend a portion of their finite privacy-loss budget to figure out how variables were coded and re-do the analysis. Annotating the dataset with a robust descriptions and context for each variable, including how it has been coded, seems critical for eliminating such confusions. 



The DP creator prototype also allowed participants to see and adjust the error due to noise on each of the statistics they were releasing.
While participants found it helpful to be able to manage error before deciding on a final allocation of the privacy-loss budget, they often found the value of the error displayed to be meaningless or confusing. For example, during one session, the prototype displayed the error for a histogram statistic to be approximately 300 counts, meaning that \DP{each bin would be off by no more than 300 observations}. However, the analyst did not know how many total observations were in the dataset (which is often considered a privacy-sensitive quantity itself) and consequently, could not decipher whether this error was \DP{overpowering the actual value of the statistic}. 
\begin{quote}
    I'm finding it hard to interpret the error\ldots it's not meaningful because I have no idea how big my dataset is. What if the dataset has 1000 versus 10000000 observations? I want to see the number of observations to understand the scale of data versus error. (A10)
\end{quote}
Thus, it may be helpful as a default to privately release summary statistics that are helpful for contextualizing other analyses, such as the number of observations in the data, even if that requires spending a portion of the privacy-loss budget upfront. 

\subsubsection{Participants assumed new risks and responsibilities regarding sensitive data.}
\label{sec:risks-responsibilities}

Several depositors and analysts said that they felt hesitant or fearful when using the tool. A few participants said that this was because of the black-box nature of the tool.
\begin{quote}
    This requires a ton of trust in the tool. It seems kind of black box-y\ldots even with the open-source code \ldots it seems a little scary to me. Linking me out to places where I can see how the black box works, that would be reassuring. (D11)
\end{quote}
However, most of the participants said that their lack of trust was more due to having to make many decisions they did not feel qualified to make as researchers, or that they did not feel that they could oversee as administrators. They said they trusted the guarantees of the algorithms within the tool, but they felt that they could inadvertently expose the sensitive data by making wrong choices. Depositors, especially, worried about being held liable for these mistakes. 
\begin{quote}
    I believe you need to be educated in DP to feel confident in the tool and for it to be useful. Uneducated researchers would not use the tool out of fear that they'd use it improperly and harm their own career (D1)
\end{quote}
\begin{quote}
    Since I don't know much about DP, I don’t feel like I could own the process. (D7)
\end{quote}
\begin{quote}
    From my perspective as an archivist, I want the researchers to know that this tool helps, but there are no guarantees. I would want to vet this with legal counsel to make sure we are not responsible if we give [researchers] this tool. And I'd remind them that there is no perfect answer to privacy. (D12)
\end{quote}
Indeed, as described in Section~\ref{sec:dp-in-practice}, there are several privacy decisions that the depositor must make within the tool before releasing statistics, such as: \DP{validating that the data is suitable for DP, answering questions about how the data was sampled, choosing the degree of sensitivity of the dataset, according to institutional guidelines (or manually choosing privacy-loss parameters, if desired), and selecting ranges and categories for variables.}
Depositors said that they would feel more comfortable making these choices if they were well-versed in DP; otherwise, they would hesitate to use the tool on their group's sensitive data in real life. Participants repeatedly asked for more guidance or validation from the tool.
\begin{quote}
    I am concerned with messing up the choices. Some way to validate would be really helpful. (D4)
\end{quote}
\begin{quote}
    If I keep clicking ``I’m unsure," will some assistance like Clippy pop up? Will [the tool] be more careful about the masking? (D13)
\end{quote}
Both depositors and analysts also reported feeling hesitant given that they only had limited trials to release or access privacy-preserving statistics. They knew that every failed attempt would use some privacy-loss budget.
\begin{quote}
    I felt very cognizant of the privacy-loss budget while exploring the data, and kept thinking: will I have enough for the analysis I want to make later? If I miss something while exploring, that will screw up my analysis\ldots this is a hard tradeoff to make (A8)
\end{quote}

\subsubsection{Participants emphasized the need for DP to be well-integrated into their data analysis pipelines.}
\label{sec:pipeline-integration}

Both depositors and analysts emphasized that for DP to be usable and useful, the user interface and workflow of DP Creator must integrate well into a researcher's data analysis pipeline. This starts with even uploading the data.
\begin{quote}
    How does data upload work? What sort of preprocessing am I allowed to do before I upload my data? (D5)
\end{quote}
In addition, the workflow and user interface should match the preferences of the analyst. One analyst with a social sciences background stated that she preferred to have a more interactive interface that includes dragging and dropping variables onto a canvas. However, another analyst with a statistics background said that he disliked the graphical user interface and would have liked to write a script instead to perform computations over the variables in the dataset. 
\begin{quote}
    I would want to see the full suite of data visualizations that I'm used to (A3)
\end{quote}
\begin{quote}
    One drawback is having to switch workflows, if I want to merge something that comes from this into another tool (A9)
\end{quote}
Multiple participants had concerns about how the statistics could be cited or published. As DP has not been widely used within research studies, the answer to these questions remains unclear.
\begin{quote}
    How would I report this? I mean, how would I use this statistic in a publication? How would I write up that I am using a privacy-preserving statistic? (A10)
\end{quote}
\begin{quote}
    What happens after I get the privacy-preserving result? Would I have to request access to the full [un-masked] statistic? How would it appear in an academic publication? Part of the point is that you don’t need access to the full data, if you have a [privacy-preserving] estimate here. How do I describe or report that I’m using differentially private statistics? (A16)
\end{quote}
In addition, several participants voiced that they would like different options for exporting the differentially private statistics outputted by the tool (eg. in a CSV format, or in a Jupyter notebook with visualizations), in order
to integrate them into various downstream analysis routines or share the results with collaborators. 


\subsection{Utility of DP}
\label{sec:utility-of-dp}

Next, we discuss themes related to the utility of differential privacy for advancing goals of open access (RQ2). Overall,
participants were optimistic about the ability of DP to make summary statistics of sensitive datasets more available to the general public.
However, participants expressed that constraints around accessing raw data and pre-planning one's analyses in order to manage the limited privacy-loss budget would make it hard to engage in exploratory data analysis. In addition, the many invisible steps within the data analysis process, along with the noise added for privacy, would make it hard for DP to facilitate replication studies. Overall, participants conveyed to us that training and expertise was needed to guide the many decisions they were required to make within the tool.

\subsubsection{Participants felt that DP would be beneficial for wider access to the public.}
\label{sec:wider-access}
Overall, participants were optimistic that DP would help members of the public have greater access to sensitive datasets. They explained that this would not only be beneficial for the lay public, but also for researchers and data stewards who would like to convey information to the public without violating privacy standards.
\begin{quote}
    This tool would be a huge benefit for the general public gaining access to sensitive data directly and not have to rely on intermediate researchers with access. (D4)
\end{quote}
\begin{quote}
    Our data archive has a number of datasets with summary information. Being able to use some broad estimates would be useful, especially for government agencies who hold the data. They'd like to have a privacy-preserving number to use publicly. (D18)
\end{quote}
In addition, several participants expressed that the level of noise added to the statistics was lower than they had expected, and that in many cases, this meant that the privacy-preserving statistics would have sufficient accuracy for simple applications.

\subsubsection{Participants felt that DP would pose challenges for exploratory data analysis.}
\label{sec:exploration}
However, participants were neutral or hesitant about the utility of DP for exploratory analysis, given the challenges of not being able to see the raw data. In particular, analysts expressed that data exploration would be challenging without access to contextual information about variables and under the constraints of a privacy-loss budget.
\begin{quote}
    It might be useful for data analysis, but only if I had labels for the variables in order to have a more complete understanding of the dataset. Ideally, I will already know my outcome before I begin. (A2)
\end{quote}
\begin{quote}
    I'd rather have fake data to get sense for how the dataset is structured, what the variables are. From analyst perspective, you don’t really know what you are looking for, you don’t know what’s useful to spend [privacy-loss budget] on. Hard to explore something with a budget, and you might run out before you figure out what you want, the objective of the exploration. (A3)
\end{quote}

\subsubsection{Participants felt that DP would not necessarily facilitate  replication of studies.}
\label{sec:replication}
Participants were also not optimistic about the potential for DP to facilitate replication of scientific studies. They explained that scientific analyses are quite complex, and even if the analyses are made publicly available, most researchers do a lot of data processing aside from the code or steps they do publish. 
\begin{quote}
    Could be feasible for summary statistics/regressions but more complicated things would be harder. A lot of times extra stuff and intermediate steps happen, like transformations, that makes it hard to replicate. (A3)
\end{quote}
\begin{quote}
    Not being able to see the actual dataset limits me. It would be hard to trust that everything is working properly. I wouldn’t be able to check the assumptions that were made even if I got similar point estimates. (A6)
\end{quote}
In addition, the constraint of only being able to access noisy estimates through DP would pose challenges for comparisons with prior results.
\begin{quote}
    Unclear how DP would confirm or deny reproducibility because of the noise added. (A19)
\end{quote}

\subsubsection{Participants felt that using DP, even with a tool designed for non-experts, requires training, guidance, and expertise.}
\label{sec:expertise}
All participant groups (in terms of roles and background) expressed that even though the DP Creator tool abstracted away many of the complexities of DP, users would still require training to be able to use DP safely and confidently. Data administrators were the most comfortable with using the tool out of the three groups, likely because of their experience making decisions about privacy and access, but administrators were concerned about giving this tool to the researchers who access their archive's data; they emphasized that researchers would need more education and training.
\begin{quote}
Researchers would need more education about what they are trying to do, what the parameters mean in practice. Overall they need more information outside of the tool, and then be able to access specific guidance within the tool. They need to know: where am I in the process? Why this leads to that, what am I doing, and how the pieces fit together, not just the what, but the WHY. (A10)
\end{quote}
However, many of our participants were optimistic that DP would be valuable for researchers once they were able to go through some training, and they provided suggestions on how to implement educational programs.
\begin{quote}
I feel like this is pretty cool; something's happening here. But need education in layman's terms. It's like seeing a doctor in a white coat. I need more about what this is trying to accomplish, and more about recommending the defaults. I'd like to see workshops that would support this type of training\ldots The payoff will be worth it for researchers. They will feel more comfortable sharing their data. (D12)
\end{quote}
In addition, a few of our participants suggested that non-experts could be more confident using DP if they had access to consultation with experts or guidance from trained IRB administrators when making privacy-sensitive decisions throughout the data analysis process. 

In summary, participants expressed that DP tools such as this one would be useful in sharing summary statistics of sensitive datasets to the public. However, participants outlined several barriers towards using DP for exploratory data analysis and replication of scientific studies, and there seems to be more work needed to make DP useful for research purposes. They suggested that using DP to share or analyze sensitive datasets requires guidance and education beyond what might be available within a user-facing tool.

\section{Discussion}
\label{sec:discussion}

In this section, we turn to the final research question (RQ3): \remove{What are the broader implications of DP for data science?} \revision{What changes need to be made in the data science workflow to address the barriers and achieve benefits (from RQ1 and RQ2) of DP?} With this question, we aim to think about DP as ``practice rather than just product"~\cite{grint1997machine}, and to interrogate the broader ethical and epistemic dimensions of privacy-preserving data science. We discuss how our work may inform the design of tools, educational resources, and governance around safe, open access to sensitive data.


\subsection{DP changes every step of the data science workflow}
\label{sec:data-science-workflows}

\begin{figure}
    \centering
    \scalebox{0.6}{
\includegraphics[width=0.8\textwidth]{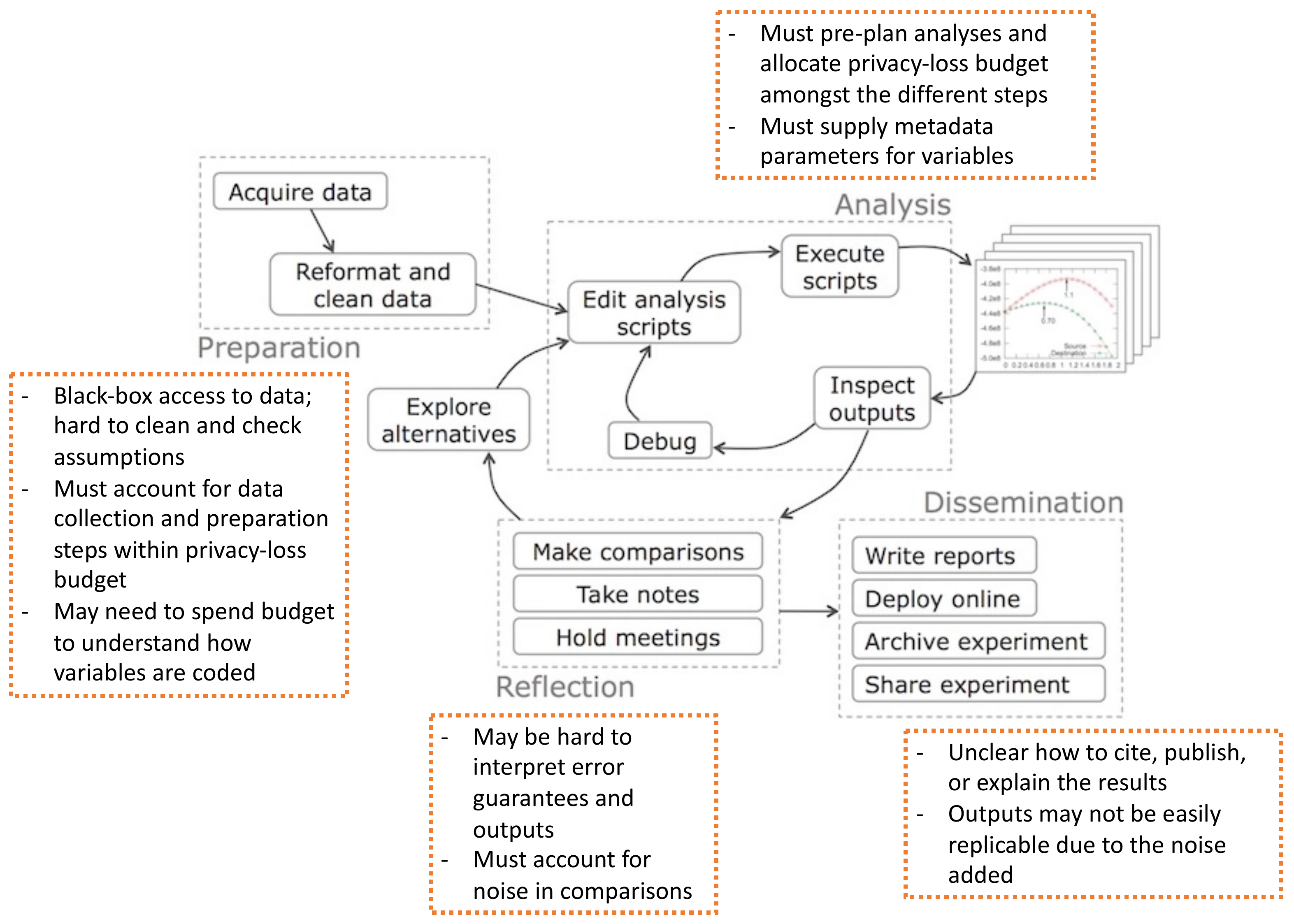}}
    \caption{The challenges posed by DP (listed in the orange, dotted boxes) for data analysts in each stage of Philip Guo's data science workflow~\cite{guo2013data}.}
    \label{fig:data-science-workflow-DP}
\Description{Philip Guo's data science workflow annotated at each stage with additional challenges posed by DP.}
\end{figure}

From our interviews with participants, it is clear that DP changes every step of the data science workflow. To examine this more clearly, we consider Philip Guo's categorization of four stages of the data science workflow~\cite{guo2013data}: \emph{preparation} of the data, iterative \emph{analysis} and \emph{reflection} to interpret the outputs, and \emph{dissemination} of results through written reports or code. Below, we describe how DP poses challenges in each stage. These challenges are summarized in Figure~\ref{fig:data-science-workflow-DP}, where we have annotated Guo's diagram with the additional considerations a data analyst must make when using DP.

\textit{Preparation.} Data scientists today spend up to 80\% of their time on tasks such as data cleaning, data filtering, and data formatting before pursuing further analysis~\cite{dasu2003exploratory}. The importance of the data cleaning process cannot be overstated: Guo writes that ``the chore of data reformatting and cleaning can lend insights into what assumptions are safe to make about the data, what idiosyncrasies exist in the collection process, and what models and analyses are appropriate to apply"~\cite{guo2013data}. However, DP creates barriers to the data preparation process. Data depositors are able to access the raw data and perform data cleaning, but DP requires that all aspects of the data preparation process -- including reformatting, imputing missing data, and cleaning up errors -- be accounted for within the privacy-loss budget, leaving less budget for the subsequent release. (Note that since it is difficult to systematically account for all the steps involved in data preparation, deployments of DP, such as at the Census Bureau, have decided to consider differential privacy only after the data have been prepared~\cite{abowd20222020}.)
Analysts face an additional challenge: they are only able to access the data through a black-box system. Although they are able to perform transformations on the data, which still need to be accounted for within the privacy-loss budget, they may need to rely heavily on whatever data preparation had been done by the depositor. Our interviews showed that without having access to the raw data, analysts struggled to check assumptions (A2, A8), figure out how data was collected and coded (A9), and leverage contextual information (A3, A10) in order to make decisions throughout the data analysis process. If this information about the dataset is not documented, the analyst may have to spend some of their privacy-loss budget to understand how the data was collected, coded and cleaned.

\textit{Analysis and Reflection.} Next, the iterative analysis and reflection stages are considered key aspects of modern data science~\cite{dasu2003exploratory}. Yet, using DP requires users to supply several metadata parameters for the analysis, and participants struggled with not knowing exactly how these parameters would affect the privacy and utility of the outputs. They had trouble interpreting the uncertainty measures (such as error for a histogram query) without having appropriate contextual information (A2, A10). Because of the high-level nature of this prototype and our limited time with each participant, they were not able to investigate the noise was added or what algorithm was used (A6), which made it hard for them to reflect on the outputs. Fortunately, DP enables all data analysis code to be publicly available in an open-source library, so with additional time, the analysts would be able to inspect the scripts that are run on the data. As the analysis has to be done under the constraints of a finite privacy-loss budget, the user is unable to run many trials of the same algorithm to see if it makes sense, or to make unlimited changes based on reflection and discussion (A3, A8). Participants expressed that these limits imposed by DP would be a major barrier to its utility.

\textit{Dissemination.} Finally, participants were unsure how one would disseminate the outputs of a DP analysis in publications or code repositories. In particular, they were not sure how to cite or explain the results (A10, A16). They were also concerned that due to the noise added to the outputs, the results would not be reproducible (A19).

Examining each stage of the workflow underscores that DP requires reconfiguring the practices of data science from start to finish. Below, we consider ethical, epistemic, and governance questions that arise when attempting to align data science with the constraints posed by DP.
\subsection{Ethical and epistemic implications for privacy-preserving data science}
\label{sec:ethical-epistemic-implications}

From our interviews, as well as our understanding of the foundations of DP, we find that adhering to the framework of DP requires a new set of technical practices, which are a logical consequence of the mathematical definition of DP. For example, it can be shown mathematically that setting hyperparameters and bounds in a data-dependent way can compromise the privacy guarantee, so these steps must be performed without relying on the raw data or by using some of the privacy-loss budget~\cite{dwork2014algorithmic}. From the perspective of adhering to DP, we can articulate a new vision of social practice around data that requires data depositors and analysts to be cautious and account for any data-dependent steps they take during the analysis process.
However, as was clear from our contextual inquiry and interviews, this vision of social practice of DP conflicts with existing norms of data analysis, which emphasize exploring and checking assumptions within data throughout the analysis process. 

Ultimately, if the goal is to satisfy the guarantees of DP, then this means that data analysts and data science itself must be reconfigured.
But in fixing the visions of social practice of DP and reconfiguring the user, we may be taking away the agency of data analysts. This raises important ethical and epistemic questions. Will DP positively constrain data analysts to the norms of ethical data science ~\cite{oberski2020differential}, by
helping them avoid issues of data peeking, data overfitting, and p-hacking? Or, might it degrade the quality and reach of social science, as analysts are hampered by the constraints imposed by DP? Above all, might it privilege technocratic modes of knowledge production, where human judgment is substituted for black-box manipulations through a tool? 

It is critical to examine these implications of DP for data science as a discipline, for equity and access, and for trust in data more broadly. \revision{In particular, we must consider the tensions across and within the different notions of utility offered by DP, as explored in Section~\ref{sec:utility-of-dp}.} Participants in our interviews expressed that  DP releases may provide significant benefits in terms of greater access to the public or to less-resourced institutions, but at the same time, participants demonstrated some ways in which DP limits the modes of data analyses that are possible. DP also makes the steps and decisions within data analyses much more visible and legible, which may contribute to better standards around documentation and reproducibility; yet, the addition of noise may complicate efforts to ascertain replicable findings. In addition, exposing all the subjective decisions involved in making statistics may expose the myriad ways in which data is socially constructed, potentially
rupturing a widespread `statistical imaginary' of data as an objective source of truth \cite{boyd2022differential}, and impacting public trust in statistics. \revision{These tensions demonstrate the complexities around creating shared tools, practices, and visions for ethical data science.}

As dynamics around privacy, utility, and open access are all impacted by the framework of DP, policy conversations around tradeoffs among these require care and clarity. Our interviews suggest that changing the expectations and norms of data science without sufficient explanation and justification will likely lead to more confusion than acceptance. \revision{Our work also highlights that further research is needed to understand how we can navigate tensions around the utility of DP for ethical data science.}

\subsection{Reconfiguring users and practices of data science}
\label{sec:suggestions-reconfiguring}

When we turn our attention from tools and theory to the practices of data science, new questions emerge. How can we reconfigure the practices of data analysis to be both more practical and ethical?
We provide suggestions for better integrating DP and data science.

\begin{enumerate}[listparindent=2em]

\item \textit{Information flow between data depositors and analysts.}
DP underscores that data cannot `speak for itself.' It requires explanation and translation from those who have been involved its collection and stewardship.
Our interviews indicated that one way to do this is through robust dataset annotation that can guide the DP analysis process. Data annotation is a way of encoding the domain knowledge of the data depositor.
Datasets that come with detailed schema, information about their context and collection process, and metadata for each variable such as ranges and categories, would directly address many of the challenges our participants faced when sharing or analyzing sensitive datasets. 

This recommendation coheres with the push for better annotation of datasets for the sake of data ethics and transparency, 
such as Gebru et al.'s work on “Datasheets for Datasets" ~\cite{gebru2021datasheets}. The authors recommend that datasets be accompanied by a datasheet that outlines its context, including motivation, composition, collection process, and recommended uses. The authors argue that annotating the dataset in this way is essential for boosting transparency and accountability, mitigating biases in models, promoting reproducibility, and guiding researchers in selecting appropriate datasets for their tasks. Transparency, reproducibility, and exploratory data analysis are all motivations of differential privacy as well, and providing information about the data domain for the purposes of DP can be seen as part of the process of annotating datasets to give users insight into limitations of the subsequent analysis. 

In addition, data depositors should consider releasing information about the dataset itself, such as the number of observations in the data, that would provide broader context to a wide range of data analyses. This information may be privacy-sensitive, so it must be released using some portion of the privacy-loss budget upfront. However, releasing such information by default may save analysts from using even more of their budget to explore these aspects of the data or correct errors due to not having this information at hand.

\item \textit{Consultation and guidance from experts in DP.} All of our participants asked for clarification or guidance from DP experts at some point during their experience using DP Creator. They wanted to understand not only how to approach making decisions, but also \textit{why} these decisions had to be made. It seems crucial to have consultation with DP experts for data practitioners to understand the implications of their choices during the process. 

In particular, participants asked for guidance on deciding whether a dataset was suitable for use with DP, understanding the implications of data sampling on the privacy guarantees, choosing and interpreting the privacy-loss parameters, and setting metadata parameters. In addition, they wanted help understanding the relationship of metadata parameters to privacy and utility, evaluating various DP algorithms for the same statistical task (e.g. linear regression), contextualizing the error of the DP release, and interpreting the outputs. 

Such guidance can come from tutorials geared towards practical aspects of DP. These materials should be made accessible to non-experts in DP from a wide range of backgrounds; for example, tutorials should contain explanations in clear prose as well as in math and code. In addition, materials should incorporate visualization tools for differential privacy~\cite{nanayakkara2022visualizing,budiu2022overlook}, which are promising for helping users to make informed choices regarding privacy and accuracy.

\item \textit{Context-specific education.}
Our findings also highlight how DP requires many context-specific decisions, suggesting that education must also be specific to context. For example, statistical agencies should provide training and examples that are relevant to the particular data domain and norms of their data users. In addition, participants suggested that they would like the opportunity to practice the DP data analysis process on public or synthetic data that is similar to the sensitive data they wish to work with.

\item \textit{Governance of privacy-loss parameters through trained privacy officers or IRBs.}
In many cases, participants voiced that they would have trouble making some decisions within the prototype, not just because they didn't understand how to do so, but also because these decisions might have implications for personal or institutional liability. A few participants suggested that trained privacy officers or Institutional Review Boards could be in charge of making or reviewing decisions around privacy-loss parameters and allocation of privacy-loss budget to different groups of researchers. Since these parties are already tasked with making or reviewing determinations around data sensitivity and actions taken to protect privacy, this would be a natural extension of their responsibilities and expertise. However, the process of training officers to understand DP-related issues and decisions requires significant training, education, and resources.

The connection with privacy law is, unfortunately, less clear. Recent work by Nissim et al.~\cite{nissim2017bridging} and Altman et al.~\cite{altman2021hybrid} explore the connections between DP and privacy regulations in the U.S. and Europe, but it will take time to understand how the law might affect decision-making around privacy-loss parameters and information leakage. Nevertheless, privacy officers and IRBs are well-trained to prescribe responsible and cautious actions even in the face of legal uncertainty.
\end{enumerate}

\revision{
\subsection{Suggestions for research and design of DP tools}
\label{sec:suggestions-design}
Finally, our study  offers suggestions for designing user interfaces for DP and points to important research agendas within the emerging field of usable DP. Since our study did not directly seek to test the design and research of DP tools, these lessons are not validated by our study design but rather suggested by our interviews and contextual inquiry.
\begin{enumerate}[listparindent=2em]
    \item \textit{Provide more explanation behind choosing parameters and making decisions.}
    As suggested by the responses from Section~\ref{sec:understanding-complexities} and~\ref{sec:expertise}, participants desired more information about not just how to set metadata and privacy-loss parameters, but why. In particular, they asked for more information about the privacy-accuracy tradeoff, implications of mis-estimating parameters for accuracy and liability, and guidance on making educated guesses for parameters. Outside of the tool, participants desired additional educational resources such as trainings and textbooks that would cater to researchers with no prior experience in DP. 
    
    In the near term, user interfaces should make sure to incorporate explanations for decision-making that are thorough yet digestible to a practitioner (eg. through pop-ups that provide help and links to external sources of information). In addition, collective investment into creating an Epsilon Registry~\cite{dwork2019differential} that provides information and justification around DP implementations would be incredibly helpful for individual practitioners trying to make these decisions within their own contexts. In the longer term, further research is required to understand how to provide effective explanations for selecting parameters---not just regarding privacy-loss parameters, which are discussed in the literature around usable DP, but also regarding all the metadata parameters and choices outlined in Section~\ref{sec:dp-in-practice}. 
    \item \textit{Create workflows around trust and safety for both data practitioners and data subjects.}
    The insights from Section~\ref{sec:risks-responsibilities} and~\ref{sec:expertise} indicate that DP tools need to prioritize features that help data practitioners trust their decisions and feel comfortable when navigating the DP data analysis process. To start, tools should be designed to enable practitioners to conduct practice trials on synthetic or non-sensitive data, so that practitioners get a sense of the constraints imposed by differential privacy. Even more support can be provided through interactive tutorials, tests of comprehension, and wikis around common mistakes. 
    
    In addition, tools should be designed to put in place conservative defaults whenever the practitioner is not sure how to proceed in order to protect the privacy of data subjects. For example, if the practitioner is not sure whether or not the dataset was sampled from a larger population, tools should assume it was not sampled as this would impose a stricter privacy-accuracy tradeoff. Finally, DP tools should, as much as possible, match the workflow (ie. graphical or script based) and setup of data analysis tools used by the specific practitioner population to which they cater. 
    \item \textit{Develop features for automated or depositor-led data contextualization.} 
    As explored in Section~\ref{sec:raw-data} and~\ref{sec:exploration}, data analysts required more information about the dataset than the DP Creator prototype currently provides in order to conduct analyses. Their responses suggest that DP tools should, by default, provide baseline synthetic data generation to help with the data preparation stage of the data science workflow (Section~\ref{sec:data-science-workflows}) when analysts do not have raw access to the data. 
    
    In the near term, a tool like DP Creator can prompt depositors to populate their initial release with privacy-preserving summary statistics that are helpful for contextualizing other analyses, such as the number of observations in the data, number of missing values per variable, histograms and CDFs on key variables. In the longer term, user interfaces would benefit from more sophisticated visualization tools that allow for checking assumptions about datasets. In addition, a critical open research question is how tools can automatically find the most pertinent aspects of a dataset to model synthetically for analysts.
\end{enumerate}
}

\section{Conclusion}

In this study, we explore how data analysts, depositors, and administrators who are not experts in DP engage with privacy-preserving data analysis. Through \remove{a }interviews with 19 data practitioners as they used the DP Creator prototype, we gained insight into perceptions, challenges, and opportunities around
differentially private data analysis. In particular, we found that practitioners face several challenges when using DP to share or access sensitive data, including limited understandings of the reasoning behind decisions, constraints around conducting analysis without access to raw data, new risks and responsibilities, and tensions with existing data analysis pipelines. In addition, our work highlights that DP may be useful for providing safe, open access to sensitive datasets for the public, but is less straightforward for research purposes such as exploratory data analysis and replication of studies. 

Our results should not be understood as broadly generalizable to all DP tools or interfaces. We study these questions with a limited set of researchers and data administrators. 
Nevertheless, our work suggests that using DP in practice clashes with every stage of the data science workflow. We theorize that using DP as defined will require reconfiguring the practices of data science. We identify ethics and governance questions that arise when attempting to socialize users around a new set of privacy constraints, and offer suggestions for better integrating DP into data science. We hope that our work opens up future inquiry into educational resources for data practitioners, tools for data annotation, parameter selection, privacy budgeting, and analysis, and modes of governance that we believe are necessary to bridge the gaps between DP and data science.

\begin{acks}
This work would not have been possible without the OpenDP team’s efforts to build an open-source DP library and user tool. We are grateful to Raman Prasad and Michael Shoemate for setting up local instances of DP Creator for our study participants, James Honaker and Jack Murtagh for guiding the initial stages of this study, and Ellen Kraffmiller, Danny Brooke, and Annie Wu for providing additional support.
Our sincere thanks to the Privacy Tools Project and Bridging Privacy working groups for feedback throughout various stages of this work, Sam Weiss Evans for pointers to relevant literature from Science \& Technology Studies, and Priyanka Nanayakkara, Emily Tseng, and Blase Ur for their thoughtful and generous feedback on earlier drafts of this paper. Finally, we are grateful to all of our study participants and the anonymous reviewers.

AH, JS, and SV were supported in part by Cooperative Agreement CB20ADR0160001 with the Census Bureau.  SV was also supported in part by a Simons Investigator Award. SS was involved in this work as a summer intern with the Privacy Tools Project, supported by the Harvard REU program. The views expressed in this paper are those of the authors and not those of the U.S. Census Bureau or any other sponsor.

\end{acks}

\bibliographystyle{ACM-Reference-Format}
\bibliography{main}

\clearpage
\appendix
\section{Screening Questions}
\label{sec:screening-questions}

\revision{
Questions for contacts within organizations:
\begin{enumerate}
    \item Does your group own a collection of data about individuals?
    \begin{enumerate}
        \item If yes, and subject to compliance with data use agreements, would your group be willing to use this dataset within a local instance of the DP Creator prototype for the purposes of the study? The data will not be exposed in the clear, nor will it be shared or published outside of this study.
        \item Please tell us more about this dataset (eg. size, sensitivity, format) so that we can ensure it is suitable for use in our study.
    \end{enumerate}
    \item Are there individuals in your group who are experienced in data analysis but who do not have prior experience with DP?
    \begin{enumerate}
        \item (If yes to question 1): To simulate a real data depositor and data analyst, we are looking for some participants that have worked closely with the dataset and others that have very little experience with the dataset. Please let us know if there are members of your group (who satisfy the above criteria as well) that fit these descriptions.
    \end{enumerate}
\end{enumerate}
Questions for potential participants:
    \begin{enumerate}
        \item Tell us more about your skills, background, and experience regarding statistical analysis.
        \item Tell us more about your familiarity with differential privacy.
        \item If your group has provided data, what is your familiarity with this dataset? What is your familiarity with the data domain?
    \end{enumerate}
}

\section{Interview Questions}
\label{sec:sample-questions}

\begin{enumerate}
    \item Tell us about your role or occupation and your day-to-day tasks.
    \begin{enumerate}
        \item What type of data do you work with?
        \item Tell us more about your familiarity with statistical analysis.
        \item Tell us more about your familiarity with differential privacy.
        \item What barriers do you currently face when sharing or accessing sensitive data?
    \end{enumerate}
    \item As you use the prototype, please talk out loud about what you are experiencing.
    \item Describe your experience using DP Creator as a (a) data depositor or (b) data analyst and your thoughts about its utility. What challenges or benefits did you perceive while using the prototype?
    \item Would you want to (a) share your sensitive datasets or (b) access private statistics about sensitive datasets using DP Creator? Why or why not?
    \item How well do you think the statistics you released using DP Creator (a) captured the main insights that can be learned from the data or (b) revealed useful insights about the data?
    \item How well did the types of statistics available to (a) release or (b) access through DP Creator (mean, histogram, etc) match the statistics you would have liked to (a) release or (b) access if privacy was not a concern? Were there any statistics missing from DP Creator that you consider essential for providing a useful release?
    \item In your opinion, how reasonable\revision{ or unreasonable} was the noise applied to the released statistics in terms of providing utility?
    \item What worked well about the prototype? What would you improve?
    \item In your opinion, what are some potential use cases for DP Creator? 
    \item Please comment on the following potential use cases, if you have not already: exploratory analysis, providing wider access to the public and enabling replication of scientific studies.
    \item What benefits and/or drawbacks does DP Creator provide you in your day-to-day role?
    \item Is there anything else you would like us to know about your experience using the prototype, or your thoughts on differentially private data analysis more broadly?

\end{enumerate}
\clearpage

\section{Codebook}
\label{sec:codebook}

\begin{table}[h!]
    \centering
    \begin{tabular}{l|l}
        \hline
        \multicolumn{2}{l}{\textbf{Theme} / Code} \\
        \hline
         \textbf{Comprehension} & \textbf{Making decisions} \\
         \hline
         Understood privacy-loss budget &  Difficult to check assumptions \\
         Grasped privacy vs. accuracy tradeoff & Did not know variable coding \\
         Did not comprehend purpose of metadata parameters & Wanted to perform sanity checks\\
         Asked for reading on parameters & Wanted to visualize data distribution\\
         Misconceptions about how parameters affect privacy & Misremembered variable details \\
         Conflated privacy and accuracy parameters & Misinterpreted error guarantee \\
         Maintained default values & Confused about scale of noise \\
         Questioned technical parameters & Satisified with outputs \\
         Asked to see algorithm code & \\
         \hline
         \textbf{Risks} & \textbf{Pipeline} \\
         \hline
         Worried about being held liable & Asked about pre-processing \\
         Lack of confidence in own choices & Wanted graphical workflow\\
         Hesitant to validate dataset & Wanted integration with scripts \\
         Unsure about making privacy decisions & Disliked switching workflows \\
         Cognizant of limited privacy-loss budget & Wanted to export statistics \\
         & Asked about citing statistics \\
         \hline
         \textbf{Public access} & \textbf{Exploration} \\
         \hline
         Beneficial for public & Lack of contextualization \\
         More efficient access & Preferred synthetic data \\
         Communicating about datasets & Need to know result instead of searching \\
         Usable for understanding data & \\
         \hline
         \textbf{Replication} & \textbf{Education} \\
         \hline
         Invisible intermediate steps & Not suitable for novices \\
         Lack of trust in findings & Need information outside of tool \\
         Noise makes it hard to confirm & Want explanations about the ``why" \\
          & Workshops and trainings \\
         \hline
    \end{tabular}
    \caption{Codebook that resulted from our thematic analysis of contextual inquiry and interviews with data practitioners.}
    \label{tab:codebook}
\end{table}

\end{document}